\documentclass[10pt, conference]{IEEEtran}
\IEEEoverridecommandlockouts
\usepackage{cite}
\usepackage{amsmath,amssymb,amsfonts}
\usepackage{algorithmic}
\usepackage{graphicx}
\usepackage{textcomp}
\usepackage{xcolor}
\usepackage{adjustbox}
\usepackage{stfloats}
\usepackage{multirow}
\usepackage{hyperref}
\def\BibTeX{{\rm B\kern-.05em{\sc i\kern-.025em b}\kern-.08em
    T\kern-.1667em\lower.7ex\hbox{E}\kern-.125emX}}

\begin{document}

\title{DeepForestSound: a multi-species automatic detector for passive acoustic monitoring in African tropical forests, a case study in Kibale National Park
}

\author{
\IEEEauthorblockN{Gabriel Dubus\IEEEauthorrefmark{1}\IEEEauthorrefmark{2},
Théau d'Audiffret\IEEEauthorrefmark{1}\IEEEauthorrefmark{2},
Claire Auger\IEEEauthorrefmark{1}\IEEEauthorrefmark{2},\IEEEauthorrefmark{4},
Rapha\"el Cornette\IEEEauthorrefmark{6},
Sylvain Haupert\IEEEauthorrefmark{5},
Innocent Kasekendi\IEEEauthorrefmark{2},\\ 
Raymond Katumba\IEEEauthorrefmark{2}, 
Hugo Magaldi\IEEEauthorrefmark{1}\IEEEauthorrefmark{2},
Lise Pernel\IEEEauthorrefmark{1},\IEEEauthorrefmark{6},
Harold Rugonge\IEEEauthorrefmark{2}, 
Jérôme Sueur\IEEEauthorrefmark{5},
John Justice Tibesigwa\IEEEauthorrefmark{3},\\
Sabrina Krief\IEEEauthorrefmark{1}\IEEEauthorrefmark{2}}
\IEEEauthorblockA{\IEEEauthorrefmark{1}
Eco-Anthropologie, Mus\'eum National d'Histoire Naturelle, UMR7206, Paris, France}
\IEEEauthorblockA{\IEEEauthorrefmark{2}
Sebitoli Chimpanzee Project, Great Ape Conservation Project, Sebitoli, Kibale National Park, Fort Portal, Uganda}
\IEEEauthorblockA{\IEEEauthorrefmark{3}
Uganda Wildlife Authority, Kampala, Uganda}
\IEEEauthorblockA{\IEEEauthorrefmark{4}
N'lab, Nitidae Association, Montpellier, France}
\IEEEauthorblockA{\IEEEauthorrefmark{5}
Centre d'Ecologie et des Sciences de la Conservation (CESCO), UMR CNRS Sorbonne Université 7204, \\ Mus\'eum national d'Histoire naturelle, Paris, France}
\IEEEauthorblockA{\IEEEauthorrefmark{6}
Institut de Systématique, Evolution, Biodiversité (ISYEB), UMR7205 Centre National de la Recherche Scientifique, \\ MNHN, SU, EPHE-PSL, UA, Paris, France}
}
\maketitle
\begin{abstract}
Passive Acoustic Monitoring (PAM) is widely used for biodiversity assessment. Its application in African tropical forests is limited by scarce annotated data, reducing the performance of general-purpose ecoacoustic models on underrepresented taxa. In this study, we introduce DeepForestSound (DFS), a multi-species automatic detection model designed for PAM in African tropical forests. DFS relies on a semi-supervised pipeline combining clustering of unannotated recordings with manual validation, followed by supervised fine-tuning of an Audio Spectrogram Transformer (AST) using low-rank adaptation, which is compared to a frozen-backbone linear baseline (DFS-Linear). The framework supports the detection of multiple taxonomic groups, including birds, primates, and elephants, from long-term acoustic recordings. DFS was trained on acoustic data collected in the Sebitoli area, in Kibale National Park, Uganda, and evaluated on an independent dataset recorded two years later at different locations within the same forest. This evaluation therefore assesses generalization across time and recording sites within a single tropical forest ecosystem. Across 8 out of 12 taxons, DFS outperforms existing automatic detection tools, particularly for non-avian taxa, achieving average AP values of 0.964 for primates and 0.961 for elephants. Results further show that LoRA-based fine-tuning substantially outperforms linear probing across taxa. Overall, these results demonstrate that task-oriented, region-specific training substantially improves detection performance in acoustically complex tropical environments, and highlight the potential of DFS as a practical tool for biodiversity monitoring and conservation in African rainforests.
\end{abstract}


\section{Introduction} 
Tropical rainforests host a large proportion of terrestrial biodiversity, yet they are among the most threatened ecosystems due to habitat loss, climate change, and increasing human pressure. Effective biodiversity monitoring in these environments is therefore a major conservation priority \cite{keck_global_2025}. However, traditional field-based survey methods are often costly, time-consuming, invasive, and limited in spatial and temporal coverage, or can be dangerous (e.g. direct observation of forest elephants),  particularly in dense and remote African rainforests \cite{zwerts_methods_2021}.

Passive Acoustic Monitoring (PAM) has emerged as a powerful alternative for large-scale and long-term biodiversity assessment \cite{sugai_terrestrial_2018}. 
By recording environmental sounds, PAM enables the detection and monitoring of vocalizing species such as birds \cite{bardeli_detecting_2010}, primates \cite{cauzinille_applying_2024}, and elephants \cite{wrege_acoustic_2017} with minimal disturbance. Acoustic data can be collected over extended periods and across large spatial extents, providing valuable insights into species presence, activity patterns, including for nocturnal species, and ecosystem dynamics \cite{sueur_ecoacoustics_2015}. Despite these advantages, the large volume of recordings and the high acoustic complexity of soundscapes pose significant challenges for data analysis \cite{stowell_computational_2021c}.

Recent advances in deep learning have significantly improved ecoacoustic analysis, outperforming traditional signal-processing methods in many tasks \cite{shiu_deep_2020b, stowell_computational_2021c}. Several large-scale models have been developed for general-purpose audio classification as well as for biodiversity applications, including the AST \cite{gong_ast_2021}, BirdNET \cite{kahl_birdnet_2021}, and Perch v2 \cite{merrienboer_perch_2026}. These models are typically trained on extensive datasets such as Xeno-Canto\footnote{http://www.xeno-canto.org}, iNaturalist \cite{chasmai_inaturalist_2025} or AudioSet \cite{gemmeke_audio_2017}, achieving state-of-the-art performance across a wide range of tasks. However, their effectiveness can be limited in certain contexts. In particular, biodiversity datasets often lack sufficient coverage in specific geographic regions or for rare species, leading to limited applicability in biodiversity monitoring \cite{michaud_acoustic_2025}. For example, iNaturalist contains far fewer recordings from African regions compared to Europe \cite{chasmai_inaturalist_2025}, and BirdNET has been shown to perform less accurately in Africa or Oceania compared to Europe or North America \cite{funosas_global_2026}. To address the scarcity of labeled data in such contexts, transfer learning using pretrained audio models has emerged as a promising strategy \cite{goodfellow_deep_2016}, allowing models to be effectively adapted to new tasks with limited annotated datasets, as is often the case in ecoacoustics \cite{ghani_global_2023a, mcginn_feature_2023}. This approach is particularly relevant in African rainforests, where high species diversity contrasts sharply with the limited availability of labeled acoustic data, posing a significant barrier to training high-performance classifiers.

In this context, and in line with the objectives of the One Forest Vision\footnote{https://www.oneforestvision.org/} initiative, which aims to monitor biodiversity using both camera traps \cite{magaldi_deepforestvision_2025}, and acoustic methods, we introduce DeepForestSound (DFS), an automatic detection and classification model trained for PAM in African tropical forests. Based on a semi-supervised workflow to efficiently generate large training datasets for species of interest, DFS has been trained through fine-tuning of AST. The proposed approach aims to provide a model capable of detecting multiple taxons from long-term acoustic recordings, including endangered birds and mammals, whose monitoring is critical for assessing their conservation status across their geographic range.

The objectives of this study are threefold: (i) to describe the framework used to train DFS, from data preparation based on semi-supervised clustering to supervised detection and classification; (ii) to evaluate its performance across a range of species of interest in Ugandan rainforest; and (iii) to assess the benefits of parameter-efficient fine-tuning for DFS using Low-Rank Adaptation (LoRA, \cite{hu_lora_2022}) compared to a frozen-backbone baseline (DFS-Linear) and to compare DFS with existing automatic detection tools commonly used in ecoacoustics. By addressing the challenges specific to tropical environments, DFS seeks to contribute a tool for biodiversity monitoring and conservation applications in Central and East African tropical forests, with current validation limited to the Kibale National Park ecosystem.

\section{Material and methods}
\paragraph*{Project ethics}
The data presented in this study were collected non invasively and required no contact with animals. All data were collected with the approval of Uganda Wildlife Authority and Uganda National Council for Science and Technology in Uganda (Kibale data, research permit COD/96/05).
\subsection{Species of interest}
The study focuses on a set of vocal species, present in Kibale National Park, Uganda \cite{wanyama_censusing_2010}. These species include birds, primates, and elephants, whose frequent and distinctive vocalizations make them ideal candidates for acoustic detection. The bird species include Grey crowned crane (\textit{Balearica regulorum}, endangered), Black-and-white-casqued hornbill (\textit{Bycanistes subcylindricus}), African emerald cuckoo (\textit{Chrysococcyx cupreus}), Great blue turaco (\textit{Corythaeola cristata}), Papyrus gonolek (\textit{Laniarius mufumbiri}, near threatened), Red-eyed dove (\textit{Streptopelia semitorquata}), Black-billed turaco (\textit{Tauraco schuettii}), and Tambourine dove (\textit{Turtur tympanistria}). The primate species include Black-and-white colobus (\textit{Colobus guereza}), Grey-cheeked mangabey (\textit{Lophocebus albigena}, vulnerable), and Chimpanzee (\textit{Pan troglodytes}, endangered). Finally, both species of African elephant, namely the forest elephant (\textit{Loxodonta cyclotis}, critically endangered), and the savanna elephant (\textit{L. africana}, endangered) are included in the study.

Among the targeted species, six of them—Grey crowned crane, Grey-cheeked mangabey, Papyrus gonolek, Chimpanzee, and forest and savanna elephant—are currently listed on the IUCN Red List\footnote{https://www.iucnredlist.org/}. Focusing on these species allows to address both methodological development and applied conservation objectives, providing a framework for monitoring taxa of high ecological and conservation importance.

Although elephants produce a broad repertoire of vocalizations, this study focuses on rumbles. These low-frequency signals, with fundamental frequencies typically between approximately 5 and 20 Hz, propagate over long distances 
and are the most commonly recorded elephant vocalizations in PAM studies \cite{geldenhuys_learning_2025}. Their high prevalence and favorable propagation characteristics make rumbles particularly well suited for acoustic-based monitoring in dense forest environments. In Kibale National Park, evidence of hybridization between forest and savanna elephants, including the presence of fertile hybrids, has been reported \cite{bonnald_phenotypical_2023}. This genetic admixture further complicates species-level discrimination, which is already challenging using vocalizations alone \cite{hedwig_does_2021}. Thus, detections are performed at the genus level.

\subsection{Data available}
Table~\ref{tab:annotation} summarizes the different datasets used in this study and their respective roles in the pipeline.
\paragraph{Sebitoli 2023}
The Sebitoli 2023 dataset was collected as part of a non-invasive wildlife monitoring program combining passive acoustic recorders and camera traps deployed in Sebitoli area, in the northern part of Kibale National Park, Uganda, by the Sebitoli Chimpanzee	Project (SCP). Although initially designed for targeted ecological studies, the recordings capture a wide range of vocalizing taxa and are therefore suitable for multi-species acoustic analyses.

Acoustic data were collected using 16 autonomous Song Meter Mini recorders (Wildlife Acoustics, USA), deployed in natural forest environments and operating without human presence. Recorders were configured to record mono audio at 44.1 kHz and 16-bit resolution. Recordings followed a temporally structured sampling scheme, covering the periods 06:00–11:00 and 16:00–19:00, resulting in 8 hours of audio per day per sampling point.


\paragraph{Publicly available datasets (PAD)}
To expand species coverage and support multi-species analyses, three publicly available datasets were included. XC, an open-access repository of bird vocalizations, providing extensive recordings of tropical and temperate species. Central African Primate Vocalization Dataset \cite{zwerts_introducing_2021a}, containing annotated recordings of multiple primate species. Congo Soundscapes, published by the Elephant Listening Project\footnote{ http://www.elephantlisteningproject.org/}, provides long-term acoustic recordings from forest habitats, including elephant rumbles.

\paragraph{Species-specific datasets}
In addition, we incorporated new datasets from Sebitoli area specific to chimpanzees and elephants to complement the field recordings. For chimpanzees, recordings with handheld devices (RHD) were collected during field direct observations using a portable Zoom$^{\text{TM}}$ recorder placed near vocalizing individuals. For elephants, audio segments were extracted from camera trap (CT) videos in which rumble vocalizations were present. 

\paragraph{Sebitoli 2025 (evaluation dataset)}
For evaluation purposes, we used passive acoustic data recorded at Sebitoli in 2025, consisting of recordings from 20 Song Meter Mini recorders (Wildlife Acoustics, USA) deployed at locations different from those of Sebitoli 2023. Recorders captured 1-minute audio segments every 15 minutes over a 6-month period from February to August 2025. Audio was recorded in mono at 48 kHz, 16-bit resolution. 

These recordings were specifically collected at new locations to enable a robust evaluation of the model, but within the same area. As a result, the evaluation primarily assesses the model’s ability to generalize across time and recording sites, while generalization to entirely different soundscapes is not addressed in this study. Cross-ecosystem generalization is therefore intentionally left out of scope and identified as a priority for future work.

\subsection{Semi-supervised clustering}
\begin{figure}[!t]
    \centering
    \includegraphics[width=0.99\columnwidth]{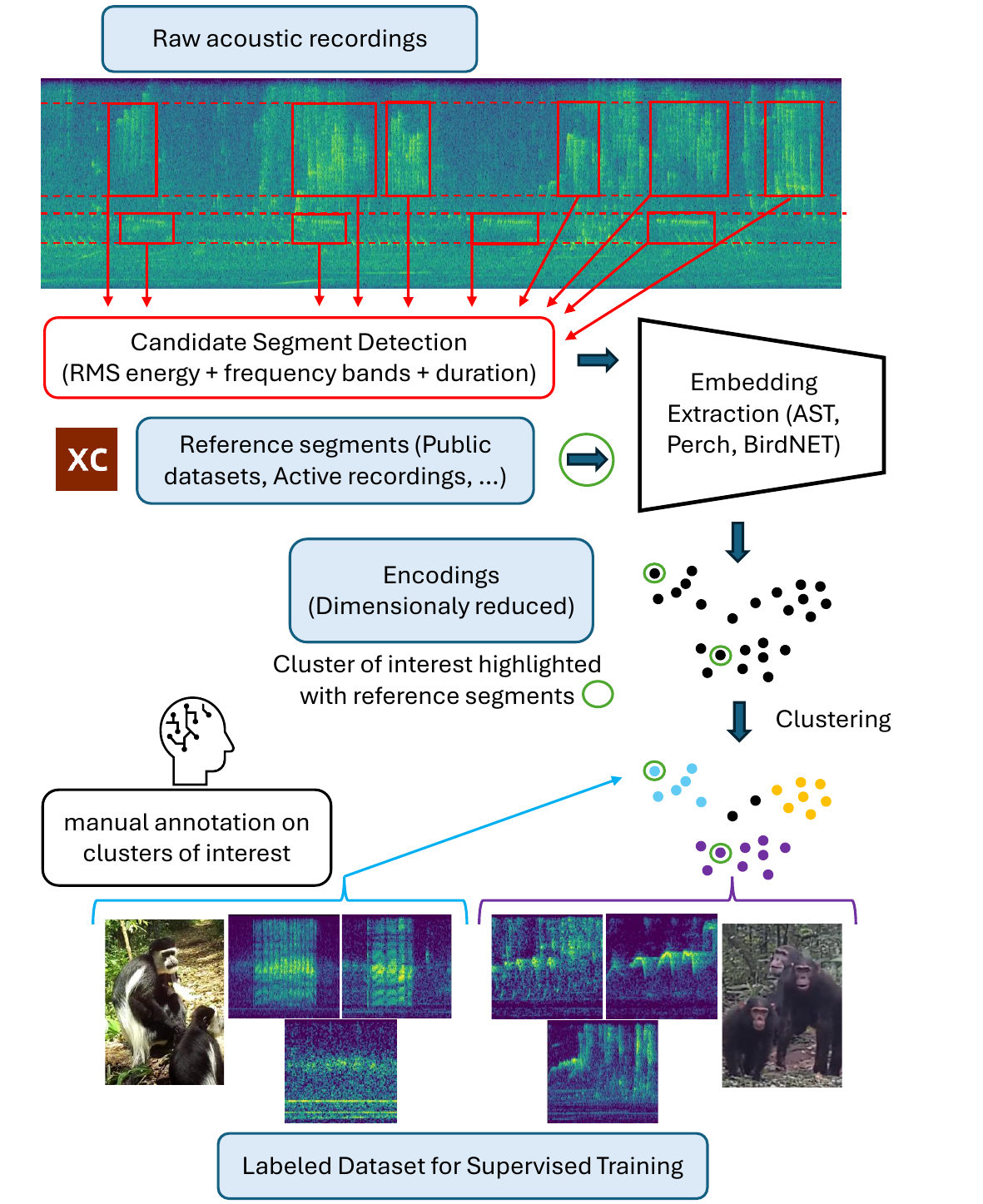}
    \caption{\small Semi-supervised clustering pipeline used. Candidate vocalizations are first detected using energy-based criteria. Then, embeddings are obtained using large pre-trained models (AST, Perch v2 or BirdNet). Embeddings of candidate vocalizations are mixed with the embeddings of annotated segments (for instance coming from XC). Dimensionality reduction and clustering, achieved with UMAP and HDBSCAN respectively, enable the extraction of species-specific vocalizations, which are manually annotated to produce a curated dataset for supervised training. Photos \copyright SCP.}
    \label{fig:schema_pipeline}
\end{figure}

The 2023 acoustic recordings were largely unannotated, which necessitated a framework to generate labeled examples suitable for training deep learning models. To address this, we developed a semi-supervised clustering pipeline to extract candidate vocalizations of species of interest, illustrated in Fig. \ref{fig:schema_pipeline}. 
In summary, the semi-supervised pipeline consists of four main steps: (i) detection of candidate vocalizations based on energy criteria, (ii) extraction of embeddings using pretrained models, (iii) clustering in a reduced feature space, and (iv) manual validation of clusters to produce labeled training data.
Similar approaches have been successfully applied in previous ecoacoustic studies to facilitate annotation in data-scarce contexts \cite{sainburg_finding_2020b, michaud_unsupervised_2023}.

First, candidate segments were detected based on dB RMS energy within literature-derived frequency bands and expected duration ranges for each species. RMS was computed over 1-second windows with 0.125-second overlap, and segments exceeding a +3 dB threshold during the specified duration were retained. Each detected segment was then encoded using embeddings from the last layer of three pretrained models used separately to capture complementary acoustic representations. AST, Perch v2 and BirdNET were used as encoders to create different feature spaces. Through manual inspection, AST was found to produce separated clusters, especially for primates, while BirdNET and Perch v2 for birds. The high-dimensional embeddings were subsequently projected to a lower-dimensional space using Uniform Manifold Approximation and Projection (UMAP) \cite{mcinnes_umap_2020} with 50 nearest neighbours and a minimum distance of 0.1, emphasizing global separation of clusters to facilitate species-level discrimination. 

Clusters were identified using Hierarchical Density-Based Spatial Clustering of Applications with Noise (HDBSCAN) \cite{mcinnes_hdbscan_2017}, with a minimum cluster size of 15, minimum samples of 2, leaf-based cluster selection, a cluster selection epsilon of 0.1, and the Euclidean distance metric. Segments classified as noise were re-clustered iteratively, applying UMAP again only on the noise points to recover smaller or less distinct vocalization groups. 

Hyperparameters for UMAP and HDBSCAN were selected empirically based on preliminary experiments. A systematic sensitivity analysis was not conducted and remains an area for future work.

To facilitate cluster inspection and semi-automatic extraction of target species, candidate segments were augmented with additional segments from publicly available and previously annotated datasets. These augmented segments were combined with random background noise from Sebitoli 2023, which ensured that clusters containing vocalizations of the species of interest were clearly separated and could be automatically extracted before manual verification.

Finally, all clusters were manually inspected to ensure high quality annotations. This process allowed the creation of a curated labeled dataset from otherwise unannotated recordings, suitable for training and evaluating supervised detection and classification models.

\subsection{Manual annotation}

Following the semi-supervised clustering stage, all candidate vocalizations were subjected to a manual annotation and validation process to ensure the quality and reliability of the training data. This step aimed to confirm species identity, remove false positives, and eliminate acoustically ambiguous segments that could degrade model performance. False positives were retained as negative segments for model training.

Each cluster extracted from the semi-supervised pipeline was inspected using spectrogram representations and audio playback. Segments were retained only if they exhibited clear and unambiguous vocal characteristics of the target species. 

Table~\ref{tab:annotation} summarizes the final number of annotated vocalizations per species and per data source after manual validation. Recordings from Sebitoli 2023 are completed by additional annotated vocalizations from multiple external sources presented earlier. The resulting dataset provides a curated collection of vocalizations spanning birds, primates, and elephants, forming a reliable basis for training the supervised detection models presented in this study.

\begin{table*}[b]
\scriptsize
\centering
\caption{Number of annotated vocalizations used for training and number of positive files used for evaluation, per taxon. Sebitoli 2023 refers to passive acoustic recordings collected in 2023 in the Sebitoli area. RHD denotes recordings collected with handheld devices during direct field observations. CT refers to audio segments extracted from camera trap videos. PAD denotes publicly available datasets. Sebitoli 2025 (eval) reports the number of positive one-minute audio files used for evaluation.}
\label{tab:annotation}
\begin{tabular}{l|ccccc|c}
\hline
\textbf{Species} 
& \textbf{Total (train)} 
& \textbf{Sebitoli 2023} 
& \textbf{RHD} 
& \textbf{CT} 
& \textbf{PAD} 
& \textbf{Sebitoli 2025 (eval)} \\ \hline
Chimpanzee                       
& \textbf{8 131} & 2 821 & 608 & 0 & 4 702 &  115 \\
Black-and-white-casqued hornbill 
& \textbf{7 942} & 7 942 & 0 & 0  & 0  & 170 \\
Black-and-white colobus          
& \textbf{6 456} & 6 456 & 0 & 0   & 0   & 116 \\
Black-billed turaco              
& \textbf{5 218} & 5 218 & 0 & 0   & 0   & 127 \\
Red-eyed dove                    
& \textbf{5 047} & 5 047 & 0 & 0   & 0   & 94 \\
Great Blue turaco                
& \textbf{3 122} & 3 122 & 0 & 0   & 0   & 102 \\
African emerald cuckoo           
& \textbf{3 067} & 2 808 & 0 & 0   & 259 & 122 \\
Grey-cheeked mangabey            
& \textbf{2 443} & 2 443 & 0 & 0    & 0   & 98 \\
Elephant (forest and savanna)                       
& \textbf{1 418} & 506  & 0 & 912  & 0   & 62 \\
Tambourine dove                  
& \textbf{1 415} & 1 279 & 0 & 0   & 136 & 236 \\
Grey crowned crane               
& \textbf{1 185} & 994  & 0 & 0   & 191 & 92 \\
Papyrus gonolek                  
& \textbf{599}   & 503  & 0 & 0    & 96  & 62 \\ \hline
\textbf{Total}                  
& \textbf{46 043} & \textbf{39 139} & \textbf{608} & \textbf{912}   & \textbf{5 384} & \textbf{1 396} \\ \hline
\end{tabular}
\end{table*}

In addition to the training dataset, a dedicated evaluation set was manually annotated to support robust performance assessment. A total of 1,200 one-minute audio files were selected from the Sebitoli 2025 dataset introduced earlier. These recordings were manually annotated using both spectrogram inspection and audio playback, enabling identification of target species vocalizations. Manual annotation was conducted by 2 trained annotators. Annotator cross-check was done to ensure consistency, and disagreements were resolved through discussion. This manual inspection of recordings required approximately 10 hours of expert work, highlighting the importance of efficient semi-supervised pipelines to scale annotation efforts.

\subsection{Audio pre-processing}

To accommodate the diverse acoustic characteristics of the target taxa, two pre-processing configurations were defined: a low-frequency (LF) configuration dedicated to elephant rumbles, and a mid-frequency (MF) configuration applied to all other species. While the two configurations differ in sampling rate and time–frequency parameterization, they share an identical post-processing and data augmentation pipeline, as described below.

All recordings were resampled to a fixed sampling rate (1 kHz for LF and 8 kHz for MF) and segmented into fixed duration excerpts of 10 seconds. When annotated vocalizations were shorter than the target duration, the vocal segment was randomly positioned within the excerpt and completed using background audio drawn from negative recordings. Equal-power crossfading was applied at segment boundaries to ensure smooth temporal transitions and avoid artificial discontinuities. Audio signals were normalized using z-score normalization prior to feature extraction.

Time–frequency representations were computed using log Mel filterbank energies. For the LF configuration, spectrograms were computed using a long analysis window (750 ms) and a frame shift of 35 ms, with 128 Mel bands spanning 3 to 250 Hz, in order to capture the low-frequency structure of elephant rumbles. For the MF configuration, a shorter analysis window (100 ms) and frame shift (19 ms) were used, with 128 Mel bands covering the 50 to 4000 Hz range, matching the spectral characteristics of bird and primate vocalizations. Resulting spectrograms were either zero-padded or truncated to a fixed length (256 samples for LF and 512 for MF), yielding uniform input dimensions. Filterbank features were subsequently normalized using global mean and variance statistics computed on the training set.

To improve robustness to acoustic variability and recording conditions, several data augmentation strategies were applied uniformly across both configurations. Time masking and frequency masking were randomly applied, following the SpecAugment paradigm \cite{park_specaugment_2019a}. In addition, for a subset of recordings acquired using handheld devices, which typically exhibit a higher signal-to-noise ratio than passive acoustic recordings, we simulated simple propagation effects characteristic of forest environments. Specifically, a frequency-dependent attenuation was applied along the frequency axis, reflecting the tendency of forest habitats to attenuate high-frequency components due to vegetation-induced scattering and absorption. In practice, a simple linear attenuation in dB as a function of frequency was applied: $A(f) = \alpha \cdot f$, following observations from physics-based sound propagation models in forest environments \cite{haupert_physicsbased_2022}. Here, $\alpha$ (in dB/kHz) was set randomly between 0.002 and 0.004, corresponding to an effective propagation distance of approximately 100-200m, based on the literature coefficient $a_0 \approx 0.019$ dB/kHz/m for tropical rainforests \cite{haupert_physicsbased_2022}. Finally, to mitigate class imbalance and increase training diversity, a mixup-based augmentation strategy was employed \cite{ferreira-paiva_survey_2022}. Positive samples were mixed with background recordings drawn from negative samples, such that the energy ratio between positive and background signals was uniformly sampled between 3 and 15 dB. The probability of applying such augmentation was adjusted according to class frequency, resulting in higher augmentation rates for underrepresented species. All features, including augmented samples, were precomputed and stored prior to training, ensuring reproducibility and reducing computational overhead during model optimization.

\subsection{Supervised detection and classification}

Supervised detection and classification were performed using an Audio Spectrogram Transformer (AST) architecture \cite{gong_ast_2021}, which has demonstrated strong performance for large-scale audio classification tasks in various fields \cite{geldenhuys_learning_2025, ferreira_transformer_2025}. Two variants of the proposed framework were implemented and evaluated. The main model, referred to as DFS, relies on parameter-efficient fine-tuning of the AST backbone using Low-Rank Adaptation (LoRA) \cite{hu_lora_2022}. A baseline variant, referred to as DFS-Linear, uses the same pretrained AST encoder but keeps all backbone parameters frozen, training only a single linear classification layer on top of the extracted embeddings.

We adopted a fine-tuning strategy based on LoRA, which injects trainable low-rank matrices into the attention layers of the pretrained model while keeping the original parameters frozen. This approach drastically reduces the number of trainable parameters (from 88.30M to 1.04M) compared to full fine-tuning or partial freezing of initial layers. Such parameter-efficient adaptation is particularly well-suited for scenarios with limited labeled data, as it enables task-specific learning with minimal computational cost.

The AST model was initialized from a version pretrained on AudioSet\cite{gemmeke_audio_2017}, providing robust generic acoustic representations. Input features were split into overlapping time-frequency patches using a stride of 10 in both dimensions, following the original AST formulation.

Two models were trained, one for LF and one for MF, using a binary cross-entropy loss function to support multi-label detection. The training set and validation set were randomly split in an 80/20 ratio. Optimization was performed using the Adam optimizer with an initial learning rate of $10^{-4}$ and a batch size of 50. No class-balanced sampling was applied during training, as class imbalance was partially addressed through the data augmentation strategies described previously.

Training was conducted for up to eight epochs on a laptop equipped with an NVIDIA RTX 3080 GPU (8 GB VRAM), with each epoch requiring approximately six hours. The training dataset comprised a total of 88,548 samples, including 26,619 negative examples and 61,929 positive samples, of which 17,431 were generated using mixup augmentation. 

\subsection{Baseline models}

To assess the performance of DFS, we compared it against three existing automatic detection models: BirdNET, Perch v2, and RDet. These models were selected to represent complementary approaches, ranging from bird-specific classifiers to large-scale multi-taxa models and a specialized detector.

\paragraph{BirdNET\cite{kahl_birdnet_2021}} is a deep learning model designed for large-scale bird sound classification. It is trained on millions of annotated bird vocalizations, primarily sourced from XC and Macaulay database\footnote{https://www.macaulaylibrary.org/}. BirdNET was evaluated only on bird species included in this study as the other targeted species are not detected. BirdNET (v2.4) was applied following the official pretrained model\footnote{https://github.com/birdnet-team/BirdNET-Analyzer} using raw audio as input: audio recordings were resampled to 48 kHz and segmented into non-overlapping 3-s windows. 

\paragraph{Perch v2\cite{merrienboer_perch_2026}} is a large-scale multi-taxa bioacoustic classifier, covering more than 15,000 species across birds, mammals, amphibians, and insects. Unlike BirdNET, Perch v2 supports non-avian taxa and is intended as a general-purpose biodiversity monitoring tool. Perch v2 was applied following the official pretrained model\footnote{https://www.kaggle.com/models/google/bird-vocalization-classifier}, using raw audio as inputs: recordings were resampled to 32 kHz and segmented into non-overlapping 5-s windows.

\paragraph{Rumble Detector (RDet)} is a specialized detector developed for the detection of forest elephant rumbles in passive acoustic recordings\footnote{https://github.com/earthtoolsmaker/forest-elephants-rumble-detection}. Following the authors recommendations, audio recordings were first downsampled to 500 Hz. Spectrograms were then computed on a logarithmic frequency scale, with amplitude expressed in dB full scale. The resulting spectrograms were provided as input to the detection model, which outputs bounding boxes corresponding to candidate rumble events in the time–frequency domain, along with associated confidence scores.

\subsection{Evaluation protocol}
Models performances were assessed using Average Precision (AP) \cite{vanmerrienboer_birds_2024}, computed as the area under the precision–recall curve, and the best $F_1$ score, defined as the harmonic mean of precision and recall. For each species, decision thresholds were independently optimized, with threshold values logarithmically varied between $10^{-7}$ and 1 \cite{funosas_global_2026}.

The best $F_1$ score was reported based on the optimal threshold, while AP was computed over all threshold values. True positives, false positives, true negatives, and false negatives were computed at the one-minute recording level.

All models were evaluated on the Sebitoli 2025 dataset, consisting of 1,200 manually annotated one-minute recordings collected at locations different from the training data. Because DFS, BirdNET, and Perch v2 operate on fixed-size audio segments, the models were applied on non-overlapping segments over each 1-minute audio file. For each species, the maximum score across all segments was retained as the final output for that species on the given file.

\section{Results}
\begin{table}[t]
\centering
\caption{Per-taxon Best $F_1$-score performance for each detector. DFS corresponds to the proposed model fine-tuned using LoRA. DFS-Linear, a baseline where the encoder is fully frozen and only a classification layer is trained. BirdNET and Perch v2 are general-purpose ecoacoustic classifiers, and RDet is a specialized elephant rumble detector.}
\label{tab:best_f1}

\begin{adjustbox}{max width=\columnwidth}
\begin{tabular}{llccccc}
\hline
 & \textbf{Taxons} & \textbf{DFS} & \textbf{DFS-Linear} & \textbf{BirdNet} & \textbf{Perch v2} & \textbf{RDet} \\
\hline

\multirow{1}{*}{Elephant} 
 & Elephant (forest and savanna) 
 & \textbf{0.901} & 0.672 & -- & 0.224 & 0.242 \\
\hline

\multirow{3}{*}{Primates}
 & Chimpanzee                 & \textbf{0.895} & 0.717 & -- & 0.439 & -- \\
 & Grey-cheeked mangabey      & \textbf{0.969} & 0.796 & -- & 0.182 & -- \\
 & Black-and-white colobus    & \textbf{0.931} & 0.733 & -- & 0.540 & -- \\
\hline

\multirow{9}{*}{Birds}
 & Grey crowned crane          & \textbf{0.967} & 0.807 & 0.541 & 0.890 & -- \\
 & Black-and-white-casqued hornbill 
                              & 0.883 & 0.860 & 0.870 & \textbf{0.914} & -- \\
 & African emerald cuckoo      & 0.729 & 0.552 & 0.739 & \textbf{0.809} & -- \\
 & Papyrus gonolek            & 0.903 & 0.267 & 0.839 & \textbf{0.953} & -- \\
 & Red-eyed dove              & \textbf{0.777} & 0.634 & 0.508 & 0.696 & -- \\
 & Tambourine dove            & 0.629 & 0.560 & 0.620 & \textbf{0.733} & -- \\
 & Great blue turaco          & \textbf{0.985} & 0.924 & 0.856 & 0.903 & -- \\
 & Black-billed turaco        & \textbf{0.972} & 0.848 & -- & 0.921 & -- \\
\hline
\end{tabular}
\end{adjustbox}
\end{table}

\begin{table}[b]
\centering
\caption{Per-taxon Average Precision (AP) performance for each detector. DFS corresponds to the proposed model fine-tuned using LoRA. DFS-Linear, a baseline where the encoder is fully frozen and only a classification layer is trained. BirdNET and Perch v2 are general-purpose ecoacoustic classifiers, and RDet is a specialized elephant rumble detector.}
\label{tab:best_AP}

\begin{adjustbox}{max width=\columnwidth}
\begin{tabular}{llccccc}
\hline
 & \textbf{Taxons} & \textbf{DFS} & \textbf{DFS-Linear} & \textbf{BirdNet} & \textbf{Perch v2} & \textbf{RDet} \\
\hline

\multirow{1}{*}{Elephant}
 & Elephant (forest and savanna)
 & \textbf{0.961} & 0.672 & -- & 0.117 & 0.101 \\
\hline

\multirow{3}{*}{Primates}
 & Chimpanzee                 & \textbf{0.934} & 0.777 & -- & 0.477 & -- \\
 & Grey-cheeked mangabey      & \textbf{0.986} & 0.838 & -- & 0.117 & -- \\
 & Black-and-white colobus    & \textbf{0.973} & 0.820 & -- & 0.515 & -- \\
\hline

\multirow{9}{*}{Birds}
 & Grey crowned crane          & \textbf{0.974} & 0.855 & 0.538 & 0.928 & -- \\
 & Black-and-white-casqued hornbill 
                              & 0.918 & 0.909 & 0.894 & \textbf{0.970} & -- \\
 & African emerald cuckoo      & 0.800 & 0.609 & 0.780 & \textbf{0.891} & -- \\
 & Papyrus gonolek            & 0.959 & 0.166 & 0.887 & \textbf{0.990} & -- \\
 & Red-eyed dove              & \textbf{0.851} & 0.649 & 0.510 & 0.751 & -- \\
 & Tambourine dove            & 0.684 & 0.606 & 0.666 & \textbf{0.806} & -- \\
 & Great blue turaco          & \textbf{0.998} & 0.969 & 0.893 & 0.948 & -- \\
 & Black-billed turaco        & \textbf{0.984} & 0.879 & -- & 0.948 & -- \\
\hline
\end{tabular}
\end{adjustbox}
\end{table}

Tables~\ref{tab:best_f1} and~\ref{tab:best_AP} summarize per-taxon detection performance in terms of best $F_1$-score and AP for DFS and baseline models.

Firstly, DFS was compared with a simplified variant, DFS-Linear, in which the AST encoder is fully frozen and only a single linear classification layer is trained. DFS-Linear consistently underperformed DFS across all taxa, with marked reductions for primates (AP from 0.964 to 0.812), birds (from 0.896 to 0.705), and elephants (from 0.961 to 0.672).

Overall, DFS achieves the highest best $F_1$-scores and AP values for the majority of evaluated species (8 out of 12), particularly for non-avian taxa (4 out of 4). Averaged across all primate species, DFS reaches a mean best $F_1$ of 0.932 and a mean AP of 0.964, while for birds it achieves an average best $F_1$ of 0.856 and an AP of 0.896. 

BirdNET achieves competitive performance on several bird species. Across avian taxa, it attains best $F_1$-scores ranging from 0.508 to 0.870 (mean best $F_1$ = 0.710) and AP values between 0.512 and 0.894 (mean AP = 0.738). 

Perch~v2 demonstrates strong performance on several bird species, achieving the highest per-species scores among baseline methods, and reaching better performance than DFS for 4 out of 8 birds. For avian taxa, Perch~v2 attains a mean best $F_1$-score of 0.852 and a mean AP of 0.904, slightly outperforming DFS on average for birds. Nevertheless, its performance on non-avian species remains substantially lower. For primates, Perch~v2 reaches a mean AP of 0.370, while for elephants it achieves 0.117, both well below the performance of DFS. 

RDet achieves limited performance on elephants (best $F_1$ = 0.242; AP = 0.101), and is not trained to detect other taxa. While its specialized design explains its taxonomic specificity, its performance remains lower than that of DFS on elephant vocalizations.

\section{Discussion}

This study presents DFS, an automatic detection and classification model designed for PAM in African tropical rainforests. DFS has been trained based on a dataset compiled using a semi-supervised framework, and supervised fine-tuning of a pretrained AST to build species-specific detectors from large volumes of passive acoustic data. Its performance was evaluated on 12 vocalizing taxa and compared with existing tools, including BirdNET, RDet, and Perch~v2.

Within an evaluation designed to assess generalization across time and recording sites, but conducted within the same area, results demonstrate that DFS achieves strong performance across taxa, particularly for non-avian species, for which general-purpose models often struggle. By leveraging task-oriented training data, domain-specific segmentation, and targeted supervision, DFS outperforms baseline methods for these taxa, highlighting the benefits of adapting detection frameworks to the acoustic characteristics and ecological context of tropical forest environments. However, the degree of acoustic similarity between sites was not explicitly quantified, which may limit the assessment of distributional shift.

A common strategy to leverage large pre-trained models is to extract feature embeddings and train a downstream classifier on frozen representations \cite{merrienboer_perch_2026, ghani_global_2023a, jana_automated_2025}. While this approach can be effective with limited labeled data, recent benchmarking has shown that embeddings without adaptation often underperform fine‑tuned models on complex classification tasks \cite{chen_no_2025}. The comparison between DFS and DFS‑Linear, in which the AST encoder is fully frozen and only a single linear classification layer is trained, reflects these limitations: DFS‑Linear exhibits consistently lower performance than DFS across taxa. Fully fine‑tuning large pretrained models can improve task performance, but in low-data regimes it is often risky due to overfitting and high computational cost \cite{cappellazzo_parameterefficient_2024}. In this context, parameter‑efficient fine‑tuning methods, such as LoRA, offer a practical alternative. The improved performance of DFS compared to DFS‑Linear suggests that lightweight backbone adaptation enables the model to capture task‑specific features, beyond what can be achieved with a fixed representation and a linear classifier. Similar observations have been reported in recent studies \cite{cappellazzo_parameterefficient_2024}, \cite{schwinger_foundation_2025}. Moreover, even if mixup-based augmentation was used to mitigate class imbalance, no ablation study was conducted to isolate its specific contribution. Therefore, performance differences for underrepresented species may result from both data scarcity and model adaptation limitations. For some underrepresented species in the training set (e.g., Papyrus gonolek with 599 samples, compared to 8,131 for Chimpanzees), the performance gap between DFS and DFS-Linear is particularly large. This observation suggests that partial adaptation of the encoder through LoRA enables the model to better adjust its feature space to minority classes, whereas a frozen representation combined with a linear classifier (DFS-Linear) may lack the flexibility required to separate rare acoustic patterns.

The model presented in this study is not intended to substitute for large-scale generalist ecoacoustic models. Perch~v2 is trained on more than 15,000 species across diverse ecosystems and is designed to offer broad taxonomic coverage, whereas DFS currently focuses on a limited set of 12 taxons selected for their ecological relevance and vocal activity in the Sebitoli forest. The objective of DFS is therefore not to outperform generalist models globally, but to complement them in targeted ecological monitoring scenarios.

This complementarity is particularly evident for primate species. While Perch~v2 achieves best scores on half of bird species (4 out of 8) without fine-tuning, its detection capability for chimpanzees and grey-cheeked mangabeys remains limited, likely due to their under-representation in large-scale training datasets. Similar trends are observed for elephant vocalizations, where DFS outperforms both the specialized RDet and generalist models. From a conservation perspective, the ability of DFS to accurately detect threatened species is relevant in the area studied. Among the 12 target taxons, six species are listed as threatened on the IUCN Red List. DFS achieves the highest detection performance for four of these taxons, and with an average best $F_1$ and AP of 0.927 and 0.963 respectively for these five taxons, the second-best model reaches 0.538 and 0.529, respectively. This illustrates its potential value for conservation-oriented monitoring programs. 

DFS is designed as an evolving model. Although the current implementation targets a limited set of African forest species, the methodology is inherently extensible. As additional annotated or semi-annotated acoustic data become available, new taxa can be incrementally integrated into the framework, at the cost of retraining the model, an operation that currently requires less than two days on a standard laptop. In terms of deployment, inference can be performed on a laptop GPU (RTX 3080), with a processing time of approximately 0.5 seconds per minute of audio (real-time factor $\approx$ 60) making DFS suitable for offline large-scale analysis of PAM datasets. 

Overall, this study highlights the importance of aligning model design with monitoring objectives. While large-scale generalist models remain indispensable for broad biodiversity assessments, task-specific frameworks such as DFS offer clear advantages for focused ecological studies, particularly for rare, threatened, or acoustically complex species. Balancing scalability and specialization will remain a key challenge for future developments in PAM, and DFS provides a practical step toward addressing this challenge in tropical forest environments.

\section{Conclusion}
We presented DFS, a task-oriented automatic detection model for PAM in Central/East African tropical forests (with validation in Kibale National Park, Uganda), trained using labeled data gathered through a semi-supervised annotation pipeline. This approach enables efficient creation of species-specific training datasets from large volumes of data.

DFS is compared with state-of-the-art detection models, while its performance on avian species is comparable to existing methods, DFS substantially outperforms generalists and specialized tools for non-avian taxa: primates and elephants. 

DFS is not intended to replace large-scale generalist ecoacoustic models, but rather to complement them in targeted ecological and conservation-oriented monitoring programs. This work highlights that, in data-scarce ecological contexts, task-specific models trained on curated local datasets can outperform large-scale generalist models. Future work will focus on extending the framework to additional species and forests, and on evaluating its transferability to other tropical soundscapes beyond the study area.

\section*{Data availability}
The audio recordings used in this study cannot be publicly shared due to legal restrictions. Processed data derived from these recordings are available from the corresponding author upon reasonable request. The code and pretrained models for DFS, along with instructions for applying the model to new recordings, are publicly available at: \url{https://github.com/MNHN-OFVI/DFS--multi-species-automatic-detector-for-pam-in-african-tropical-forests}.

\section*{Acknowledgements}
The authors are grateful to the Muséum national d’Histoire naturelle (ATM 2022 grant), the Fondation pour la Nature et l’Homme, Fondation Prince Albert 2 de Monaco, and the Fonds Français pour l’Environnement Mondial, for the long term fundings of Sebitoli Chimpanzee Project team, and to the One Forest Vision initiative, funded by the French Ministry of Higher Education and Scientific Research and the Ministry of Europe and Foreign Affairs for their support. SCP is indebted to the Ugandan field assistants for their contribution in setting up the audio-recorders and collecting data.


\sloppy
\bibliographystyle{IEEEtran}
\bibliography{dfs_refs}
\fussy

\end{document}